\documentclass{PoS}

\usepackage{amsbsy,amsmath,amsfonts,amssymb}

\def\Journal#1#2#3#4{{#1} {\bf #2} (#4) #3}

\def\NIMA{{\em Nucl.~Instrum.~Methods}~A}

\def\NPB{{\em Nucl.~Phys.}~B}
\def\PLB{{\em Phys.~Lett.}~B}
\def\PR{\em Phys.~Rev.}
\def\PRL{\em Phys.~Rev.~Lett.}
\def\PRD{{\em Phys.~Rev.}~D}
\def\ZPC{{\em Z.~Phys.}~C}
\def\EPJ{{\em Eur.~Phys.~J.}~C}

\newcommand{\Br}{{\text{Br}}}

\newcommand{\pid}{\pi^0_D}

\newcommand{\kpmmunu}{K^\pm \to \mu^\pm \nu_\mu}

\newcommand{\kpmpipizpiz}{K^\pm \to \pi^\pm \pi^0 \pi^0}

\newcommand{\kpmpipiz}{K^\pm \to \pi^\pm \pi^0}

\newcommand{\kpmpipid}{K^\pm \to \pi^\pm \pid}

\newcommand{\kpmpipig}{K^\pm \to \pi^\pm \pi^0  \gamma}

\newcommand{\kpmpigg}{K^\pm \to \pi^\pm \gamma  \gamma}

\newcommand{\kpmpieeg}{K^\pm \to \pi^\pm e^+ e^- \gamma}

\newcommand{\bdm}{\begin{displaymath}}
\newcommand{\edm}{\end{displaymath}}
\newcommand{\be}{\begin{equation}}
\newcommand{\ee}{\end{equation}}

\title{Radiative $K^\pm$ Decays from NA48/2}

\ShortTitle{Radiative $K^\pm$ Decays from NA48/2}

\author{\speaker{Rainer Wanke}\thanks{Supported by the German Federal Minister for Research and Technology under contract 05HK1UM1/1}\\
        \vspace*{2mm}
        (for the NA48/2 Collaboration) \\
        Institut f\"ur Physik, Universit\"at Mainz\\
        E-mail: \email{Rainer.Wanke@uni-mainz.de}}

\abstract{New results on radiative $K^\pm$ decays from the NA48/2 experiment are presented. 
          In the channel $\kpmpipig$ more than 1 million decays were reconstructed, leading to
          the first measurement of the interference between direct photon emission and
          inner bremsstrahlung and stringent limits on CP violation in this decay.
          For $\kpmpigg$, a precise measurement of the branching fraction was performed, 
          based on more than 1000 events.
          In addition, the related decay $\kpmpieeg$ was observed for the first time and
          measurements of the decay rate and the decay parameter $\hat{c}$ were carried out.}

\FullConference{2009 KAON International Conference KAON09,\\
		June 09 - 12 2009\\
		Tsukuba, Japan}

\begin{document}

\section{Introduction}

Radiative kaon decays offer a unique possibility to study Chiral Perturbation Theory (ChPT) in detail.
In particular, direct photon emissions as in $\kpmpipig$ decays or in decays with vanishing ${\cal O}(p^4)$
as $K^\pm \to \pi^\pm \gamma \gamma^{(\star)}$ are of theoretical interest.

The NA48/2 experiment has collected data on charged kaon decays in the years 2003 and 2004. 
The kaon beams had a momentum of 60~GeV/$c$ with $K^+$ and $K^-$ decays being recorded simultaneously,
to systematic effects in CP violation measurements~\cite{bib:kpipipi}.
The data were recorded with both a highly efficient 3-track-trigger
 for decays of charged kaons into three charged particles,
and a 1-track-trigger, which required a minimum invariant mass of the neutral decay particles to exclude
the abundant $\kpmpipiz$ and $\kpmmunu$ decays. In total, several billions of reconstructed decays were recorded.

The NA48 detector is described in detail elsewhere~\cite{bib:detector}. The main detector components
were a magnetic spectrometer, consisting of two sets of two drift chambers before and after a dipole magnet,
providing a momentum resolution of about $1.4\%$ for 20~GeV/$c$ charged tracks, and a liquid-krypton 
electromagnetic calorimeter (LKr) with an energy resolution of about $1\%$ for 20~GeV photons and electrons.

\section{$\kpmpipig$ Decays}

The total amplitude of the $\kpmpipig$ decay is the sum of two terms: inner bremsstrahlung (IB), with
the photon being emitted from the outgoing charged pion, and direct emission (DE),
where the photon is emitted from the weak vertex.
The IB component can be predicted from QED corrections to $\kpmpipiz$ in a straight-forward way~\cite{bib:ib1,bib:ib2}.
For the DE term, several studies within the framework of Chiral Perturbation Theory (ChPT) 
exist~\cite{bib:cheng1,bib:cheng2,bib:ecker,bib:ecker1,bib:ecker2}.
At ${\cal O}(p^4)$ ChPT, direct photon emission can occur through both electric ($X_E$) and 
magnetic ($X_M$) dipole transitions.
The magnetic part is the sum of a reducible amplitude, that can be calculated 
using the Wess-Zumino-Witten functional~\cite{bib:wzw1,bib:wzw2}, and a direct amplitude, which size is expected to be small.
For the electric transition no definite prediction exists.

The total decay rate as a function of the kinematic variable 
$W^2 = ( p_\pi \cdot p_\gamma )( p_K \cdot p_\gamma ) / ( m_K^2 m_\pi^2 )$
is given by
\begin{equation}
\frac{\partial \Gamma^\pm}{\partial W} = 
   \frac{\partial \Gamma^\pm_\text{\bfseries IB}}{\partial W} \left[ 1 + 
   2 \cos \left( \pm \phi + \delta_1^1 - \delta_0^2 \right) |X_E| W^2 +  
   m_\pi^4 m_K^4 \left( |X_E|^2 + |X_M|^2 \right) W^4 \right].
\end{equation}
In addition to the IB and DE contributions, the decay rate 
contains also the interference (INT) between IB and DE, 
which, apart of the strong $\pi \pi$ re-scattering phases $\delta_1^1$ and $\delta_0^2$, depends only on $X_E$
and a possible CP violating phase $\phi$. 
By measuring the INT term it is possible to disentangle the 
electric and magnetic amplitudes and to investigate possible CP violation in $\kpmpipig$.

Previous measurements have been performed by several experiments. 
The combined DE branching fraction, based on the world total of about 30000 $\kpmpipig$ events, is 
$\Br(\text{DE}) = (4.3 \pm 0.7) \times 10^{-6}$~\cite{bib:pdg08}, with the assumption of no interference term, consistent
with the only previous measurement of 
$\text{Frac}(\text{INT}) \equiv \Br(\text{INT})/\Br(\text{IB}) = (-0.4 \pm 1.6 )\%$ 
by the E787 experiment~\cite{bib:e787}. 
All previous measurements were performed in the restricted kinematic region $55 < T^\star_\pi < 90$~MeV
of the pion kinetic energy $T^\star_\pi$ in the kaon rest frame. 

NA48/2 is the first experiment which can use both $K^+$ and $K^-$ events. 
In addition, a strong suppression of $\kpmpipizpiz$ events, based on the excellent performance of the LKr calorimeter,
was implemented. This allowed to extend the kinematic region to $0 < T^\star_\pi < 80$~MeV, with a slightly stronger
upper cut due to the on-line trigger rejection of $\kpmpipiz$ events.
The remaining background, coming mainly from $\kpmpipizpiz$, was estimated with Monte Carlo simulated events 
to be less than $1\%$ of the DE contribution.
The probability of mis-identifying the odd photon was estimated to be less than $10^{-3}$.

\begin{figure}[thb]
  \mbox{\includegraphics[width=0.33\textwidth]{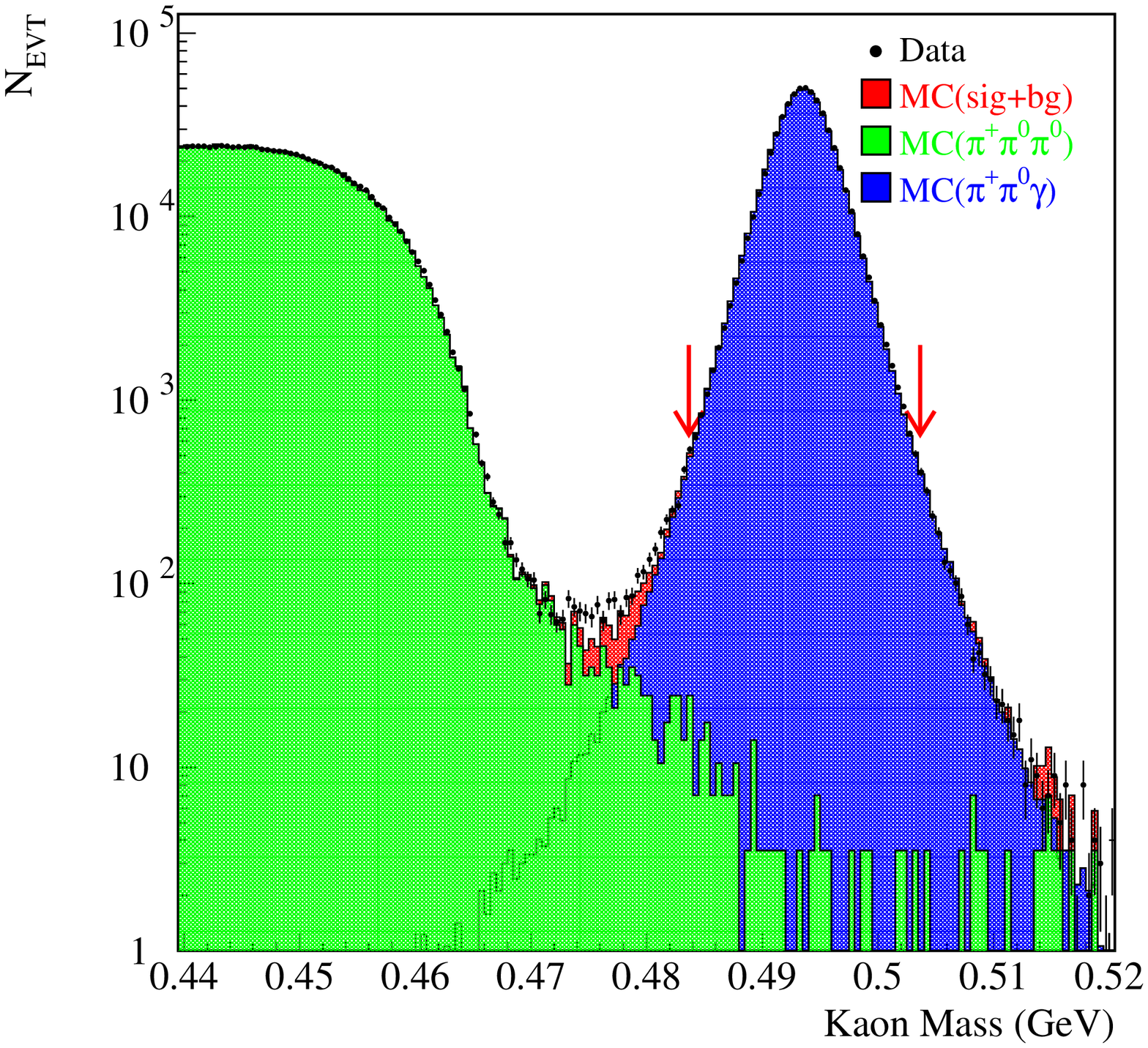}
        \hspace*{0.027\textwidth}
        \includegraphics[width=0.30\textwidth]{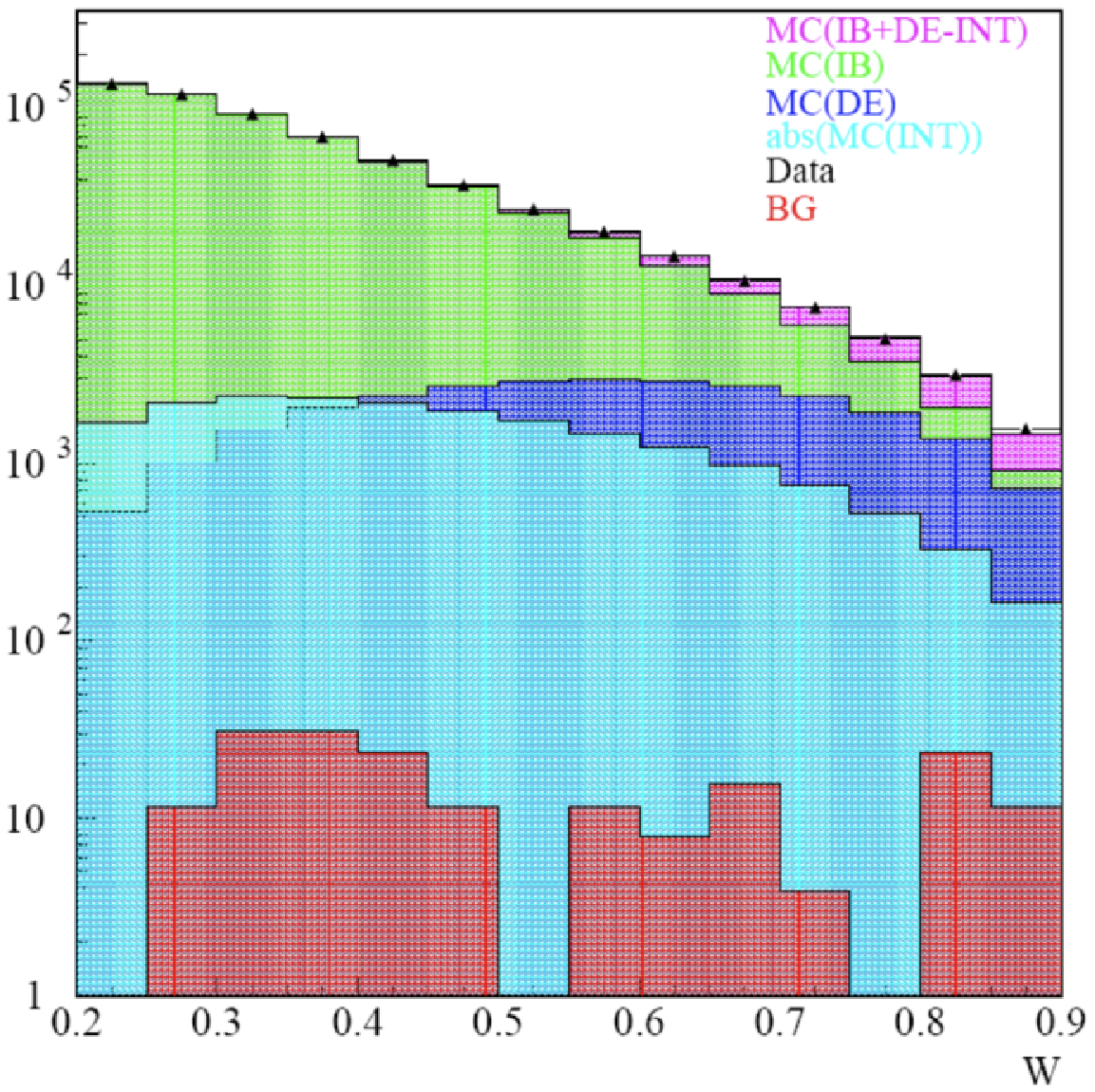}
        \hspace*{0.013\textwidth}
        \includegraphics[width=0.33\textwidth]{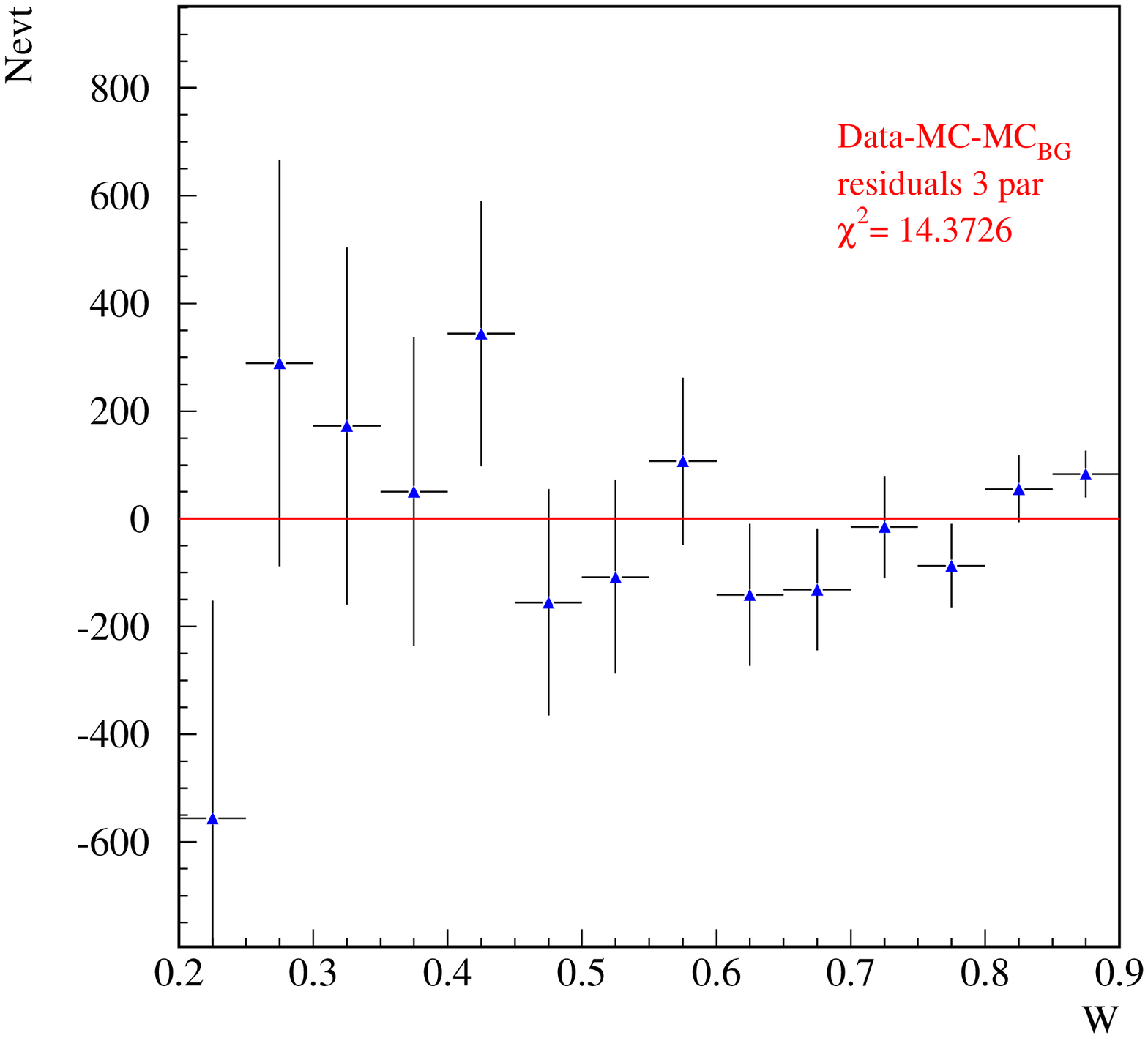}}
  \caption{{\em Left:} Selected $\kpmpipig$ candidates.
           {\em Center:} Maximum-likelihood fit of the $W$ distribution of the selected $\kpmpipig$ candidates.
           {\em Right:} Fit residuals.}
  \label{fig:Kpipig}
\end{figure}

In total, about 1 million of $\kpmpipig$ events were reconstructed by NA48/2 (Fig.~\ref{fig:Kpipig} (left)).
The extraction of the IB, DE, and INT contributions was done with an extended maximum-likelihood fit of the 
Monte Carlo $W$ distributions of the single components to the data distribution.
For the fit, the gamma energy was required to be above 5~GeV 
to be insensitive of inefficiencies of the L1 trigger for small cluster energies.
In addition, the kinematic range was restricted to $0.2 < W < 0.9$, leaving about $600\,000$ events for the fit.
The fit to the data is shown in Fig.~\ref{fig:Kpipig} and yielded
$\text{Frac}(\text{DE}) = (3.32 \pm 0.15 )\%$ and 
$\text{Frac}(\text{INT}) = (-2.35 \pm 0.35 )\%$ (for $0 < T^\star_\pi < 80$~MeV).

\begin{figure}[thb]
  \hspace*{0.14\textwidth}
  \includegraphics[width=.33\textwidth]{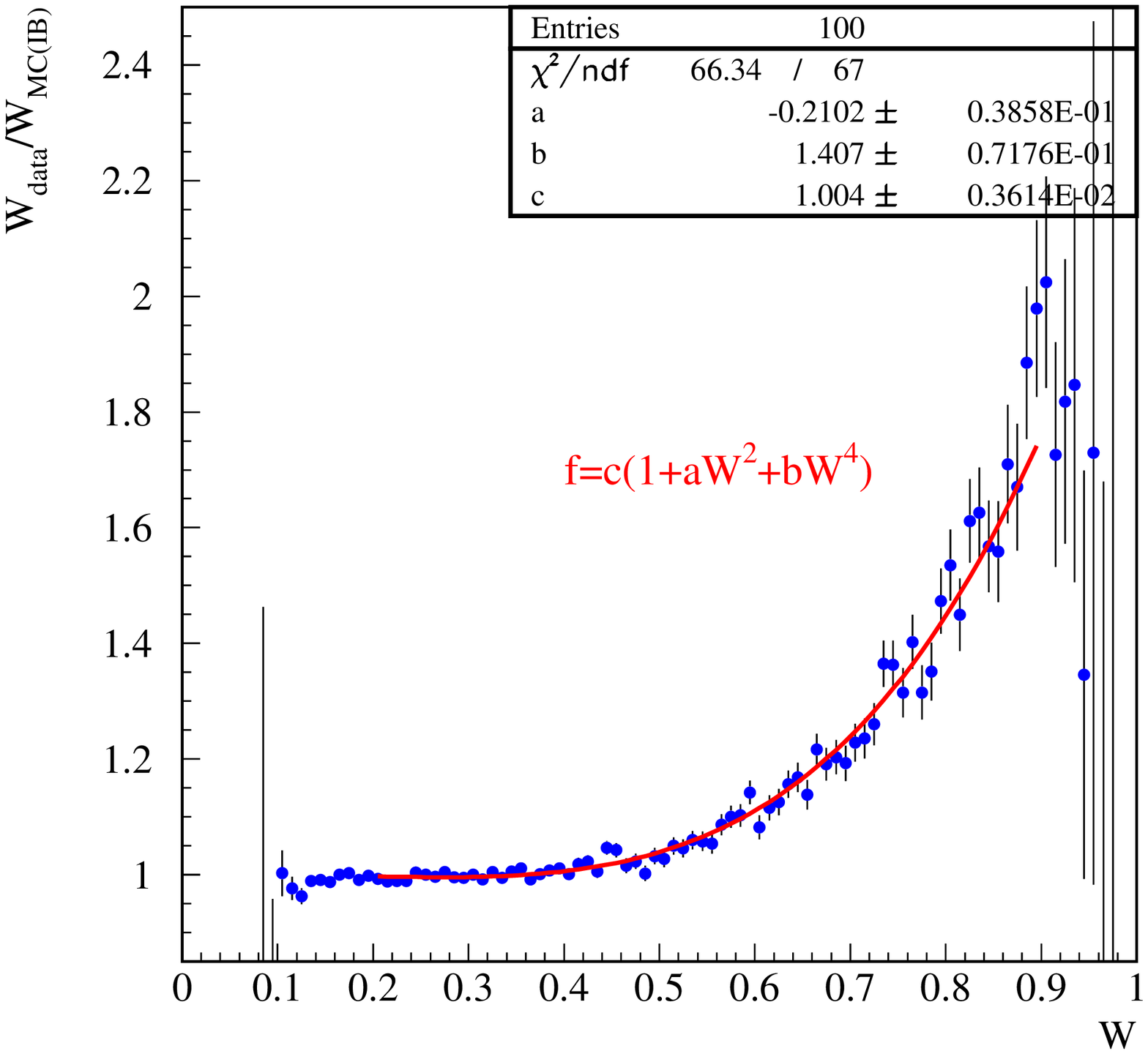}
  \hspace*{0.06\textwidth}
  \includegraphics[width=.33\textwidth]{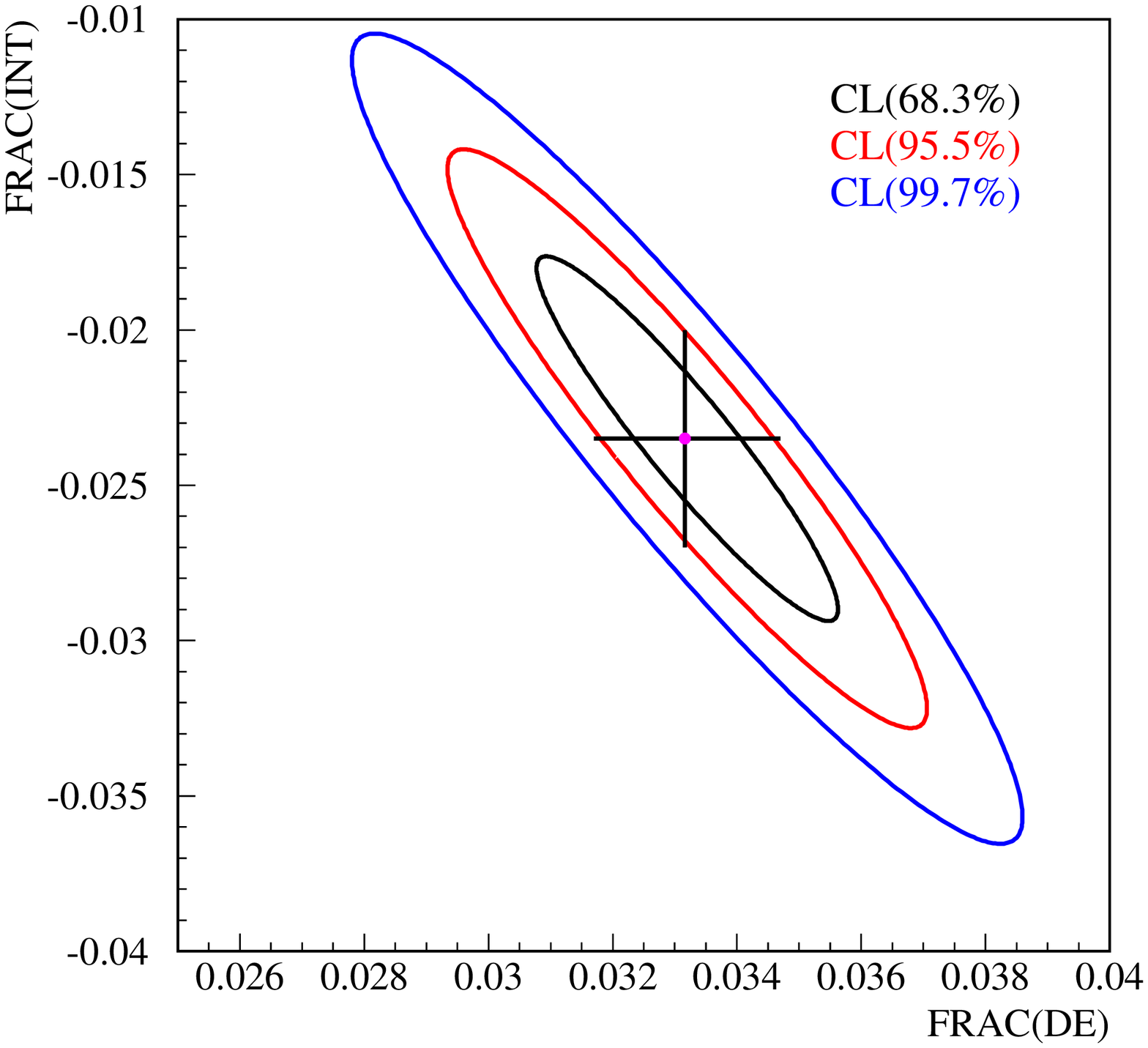}
  \caption{{\em Left:} Polynomial fit to the ratio of data over IB Monte Carlo.
           {\em Right:} Contour plot for the DE and INT terms. The black cross shows the $1\sigma$ statistical uncertainties
           of the projections.}
  \label{fig:Kpipig_fitpoly}
\end{figure}

As a cross-check, a simple polynomial fit to the data $W$ distribution, divided by the Monte Carlo IB distribution
was performed (Fig.~\ref{fig:Kpipig_fitpoly} (left)). 
Although this method does not fully correctly take into account the acceptances, the result was in perfect agreement
with the maximum-likelihood method.

Many possible systematic uncertainties were investigated. Most contributing were the description of the detector
acceptance, trigger efficiencies, and the LKr calorimeter energy scale.
The final result, including also systematic uncertainties, is
\begin{eqnarray}
  \text{Frac(DE)}_{0 < T^\star_\pi < 80 \: \text{MeV}} & = & ( \; \; \, 3.32 \pm 0.15_\text{stat} \pm 0.14_\text{syst}) \times 10^{-2}, \\
  \text{Frac(INT)}_{0 < T^\star_\pi < 80 \: \text{MeV}}  \! & = & (-2.35 \pm 0.35_\text{stat} \pm 0.39_\text{syst}) \times 10^{-2},
\end{eqnarray}
with a correlation coefficient of $-0.93$ between both values.
Fig.~\ref{fig:Kpipig_fitpoly} (right) shows the confidence regions for the statistical uncertainties.

From this, the electric and magnetic amplitudes can be extracted to
\begin{eqnarray}
  X_E & = & ( -24 \pm  4_\text{stat} \pm  4_\text{syst}) \; \text{GeV}^{-4}, \\
  X_M & = & ( 254 \pm 11_\text{stat} \pm 11_\text{syst}) \; \text{GeV}^{-4},
\end{eqnarray}
with the magnetic amplitude being very close to the WZW prediction of about $271$~GeV$^{-4}$~\cite{bib:ecker1,bib:wzwpred}.

For comparison with previous experiments, a fit with the INT term set to 0 was performed. The result, extrapolated
to the kinematic range $55 < T^\star_\pi < 90$~MeV, was
\begin{equation}
  \text{Br(DE)}^{\text{INT}=0}_{55 < T^\star_\pi < 90 \: \text{MeV}} 
                         = (2.32 \pm 0.05_\text{stat} \pm 0.08_\text{syst}) \times 10^{-6},
\end{equation}
in clear disagreement with the previous measurements. 
The $\chi^2$ of this fit was $51.0/12$ (compared to $14.3/11$ when including the INT term as a free fit parameter), 
strongly indicating the need of the INT term for a proper description of the data.

Finally, possible direct CP violation in this channel was investigated. CP violation would manifest itself in a
 decay rate asymmetry of $K^+$ with respect to $K^-$ decays and/or in different $W$ distributions for $K^+$ and $K^-$,
due to a non-vanishing phase $\phi$ in the differential decay rate.
A possible decay rate asymmetry can be expressed in an asymmetry of the total number of events, defined as
$A_N = (N_+ - R \, N_-)/(N_+ + R \, N_-)$, with $N_+$ and $N_-$ the numbers of $K^+$ and $K^-$ decays, and $R$ the ratio
of $K^+$ to $K^-$ in the beam, determined from $\kpmpipizpiz$ 
decays~\footnote{This assumes negligible CP violation in $\kpmpipizpiz$, which is consistent with the NA48/2 limit
on CP violation in the $\kpmpipizpiz$ Dalitz plot~\cite{bib:kpipipi}.}.
Removing the cuts on the $W$ range and the photon energy, thus using the complete data set of more than a million decays,
NA48/2 found $A_N = ( 0.0 \pm  1.0_\text{stat} \pm 0.6_\text{syst}) \times 10^{-3}$, corresponding to
$|A_N| < 1.5 \times 10^{-3}$ at a confidence level of $90\%$.
Extraction of the CP violating phase $\phi$ yielded $\sin \phi = -0.01 \pm 0.43$, 
equivalent to $|\sin \phi| < 0.56$ at $90\%$~CL.

Assuming the interference term to be the origin of possible CP violation, a fit to the ratio of 
the $W$ spectra of $K^+$ and $K^-$, given by
$\frac{d \Gamma^\pm}{d W} = \frac{d \Gamma^\pm_\text{IB}}{d W}  \left( 1 + (a \pm e) W^2 + b W^4 \right)$, was performed.
With the parameters $a$ and $b$ from the DE and INT fractions, a single parameter fit obtained
$A_W = e \int (\text{INT}/\text{IB}) = (-0.6 \pm 1.0) \times 10^{-3}$, in good agreement with the value of $A_N$.

\section{$\kpmpigg$ Decays}

The $\kpmpigg$ decay is of high interest in ChPT, since contributions of ${\cal O}(p^2)$ vanish, 
thus giving high sensitivity to ${\cal O}(p^4)$ and ${\cal O}(p^6)$.
The differential decay rate of $\kpmpigg$ is given as
\begin{equation}
\frac{\partial^2 \Gamma}{\partial y \partial z} = \frac{m_K}{2^9 \pi^3}
              \left[ z^2 \left( |A + B|^2 + |C|^2 \right)
                     + \left( y^2 -\frac{\scriptstyle 1}{\scriptstyle 4} \lambda(1,r_\pi^2,z) \right)^2
                     \left( |B|^2 + |D|^2 \right) \right],
\end{equation}
with $y = (E^\star_{\gamma_1} - E^\star_{\gamma_2})/m_K$ and $z = m_{\gamma\gamma}^2/m_K^2$.

At ${\cal O}(p^4)$, predominantly loop diagrams contribute, 
leading to a distinct cusp in the invariant $\gamma \gamma$ mass at twice the $\pi^+$ mass~\cite{bib:ecker}.
The amplitude $A$ is known up to a parameter $\hat{c}$, which needs to measured from experiment.
Also at ${\cal O}(p^4)$, poles and tadpole diagrams contribute to the $C$ amplitude~\cite{bib:gerard}.
At ${\cal O}(p^6)$, unitarity corrections could alter the branching fraction by $30-40\%$~\cite{bib:dambrosio}.

\begin{figure}[thb]
  \hspace*{0.04\textwidth}
  \includegraphics[width=0.45\textwidth]{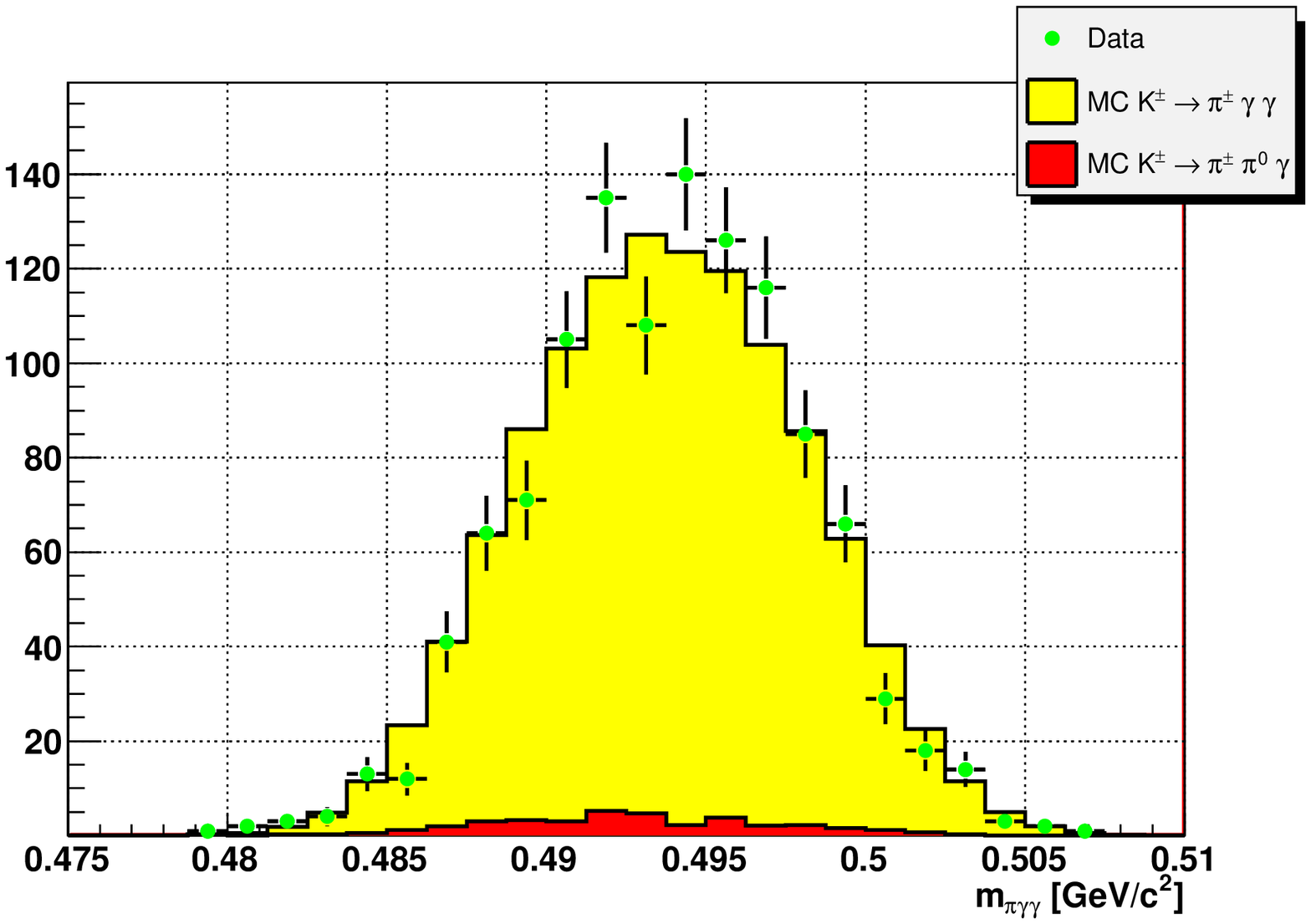}
  \hspace*{0.02\textwidth}
  \includegraphics[width=0.45\textwidth]{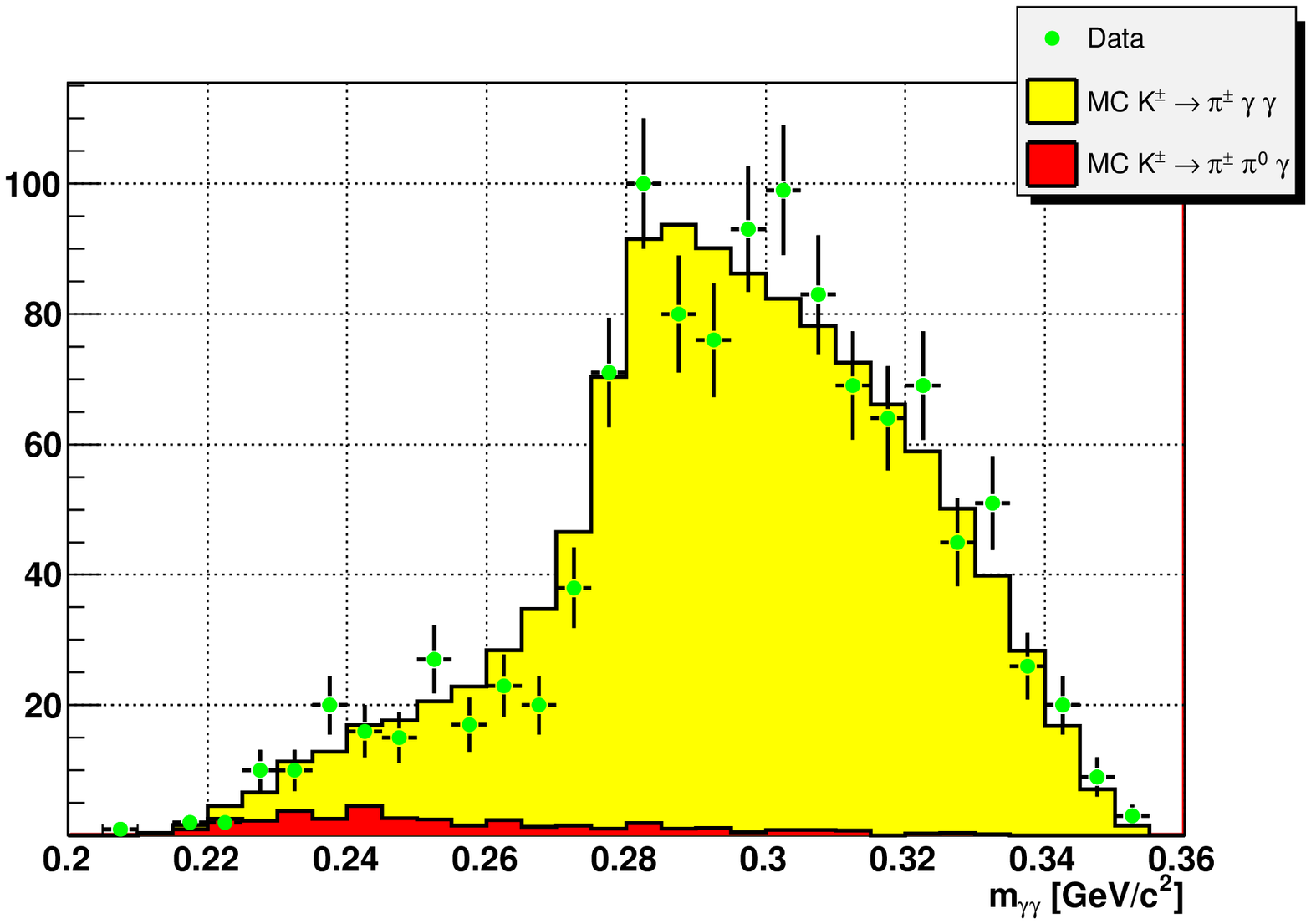}
  \caption{Selected $\kpmpigg$ candidates.
           Left: Invariant $\pi^\pm \gamma \gamma$ mass. 
           Right: Invariant $\gamma \gamma$ mass.}
  \label{fig:Kpigg}
\end{figure}

So far analyzed were about $40\%$ of the complete data set. Due to the similarity in topology to
$\kpmpipiz$ events, which were trigger suppressed, the signal trigger efficiency
was only about $40\%$. 
In total 1164 $\kpmpigg$ candidates were reconstructed and passing the selection, corresponding to about 40 times
the previous world sample.
The background contribution, mainly from $\kpmpipig$ events, was determined from Monte Carlo simulation to $3.3\%$.
The invariant $\pi^\pm \gamma \gamma$ and $\gamma \gamma$ mass distributions are shown in Fig.~\ref{fig:Kpigg}, 
the latter exhibiting the expected cusp at twice the pion mass.

Obtaining the detector acceptance from a simulation using ${\cal O}(p^6)$ ChPT with $\hat{c}=2$,
a preliminary, model-dependent branching fraction was obtained:
\begin{equation}
  \Br(\kpmpigg)_{\hat{c}=2, {\cal O}(p^6)} = (1.07 \pm 0.04_\text{stat} \pm 0.08_\text{syst}) \times 10^{-6}
\end{equation}
The systematic uncertainty is dominated by the trigger efficiency.
A model-independent measurement and the extraction of the parameter $\hat{c}$ are in preparation.

\section{$\kpmpieeg$ Decays}

The decay $\kpmpieeg$ is similar to $\kpmpigg$, with one of the photons internally converting into a pair of electrons.
As for $\kpmpigg$, in ${\cal O}(p^4)$ ChPT the branching fraction and the $e^+ e^- \gamma$ spectrum 
are determined by a single parameter $\hat{c}$. At ${\cal O}(p^6)$, 
unitarity corrections may alter the branching fraction by up to $40\%$~\cite{bib:gabbiani}.

\begin{figure}[thb]
  \hspace*{0.04\textwidth}
  \includegraphics[width=.45\textwidth]{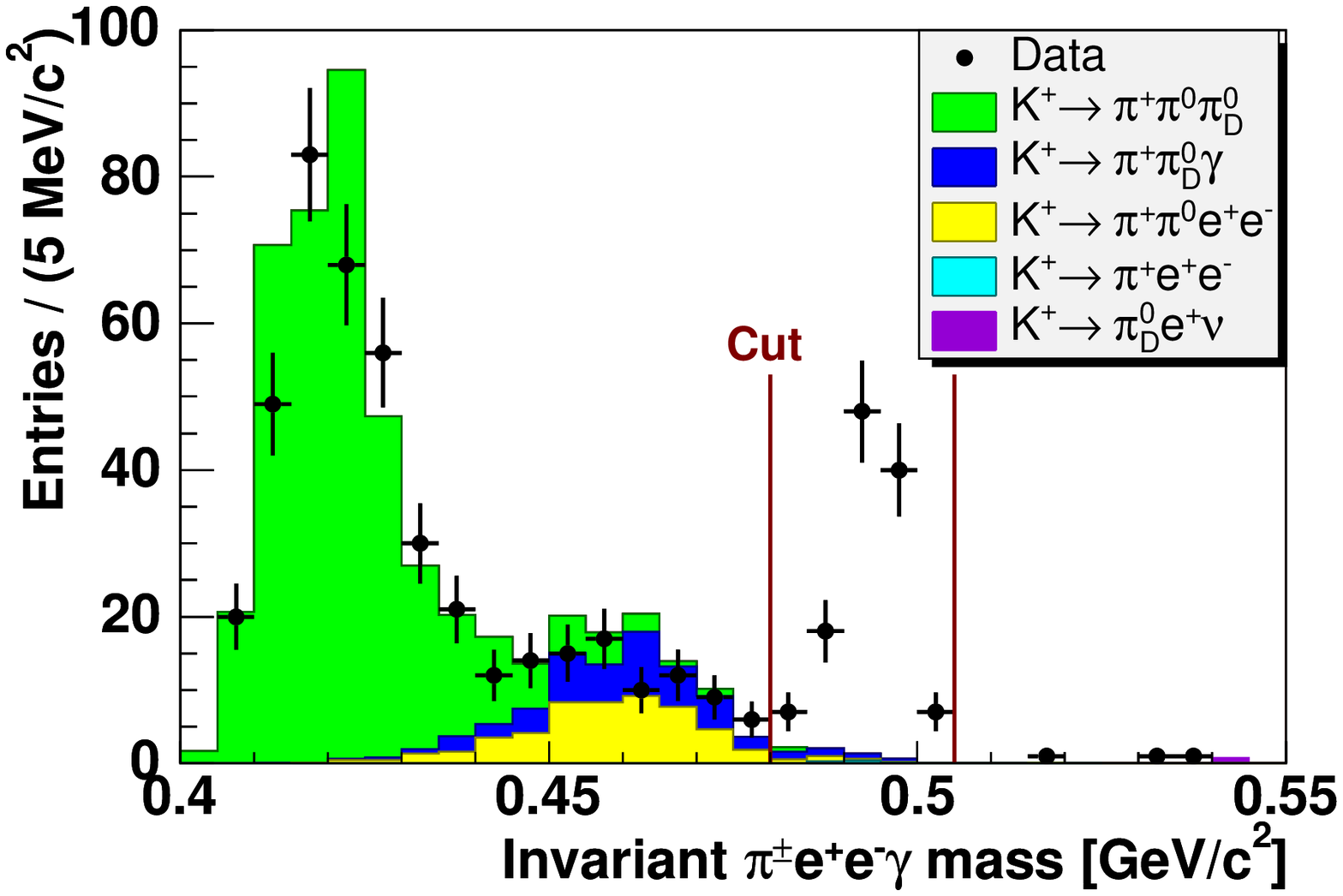}
  \hspace*{0.02\textwidth}
  \includegraphics[width=.45\textwidth]{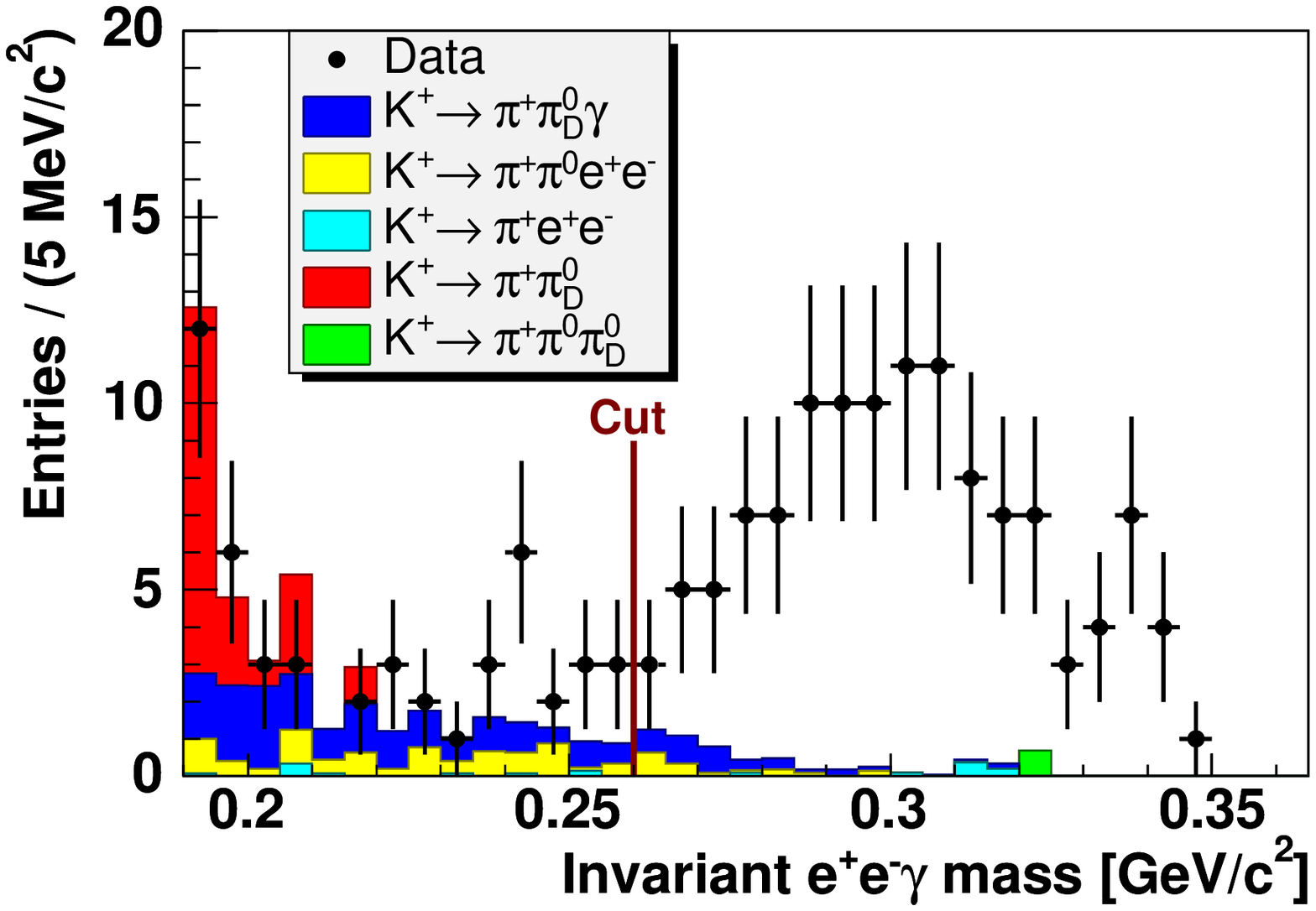}
  \caption{Invariant $\pi^+ e^+ e^- \gamma$ (left) and $e^+ e^- \gamma$ mass
           for the selected $K^\pm \to \pi^+ e^+ e^- \gamma$ candidates.}
  \label{fig:Kpieeg}
\end{figure}

Using the whole NA48/2 data set, 120 signal candidates with only small background conta\-mination were found 
(Fig.~\ref{fig:Kpieeg}). This is the first observation of this decay.

As normalization channel the abundant decay $\kpmpipid$ with $\pid \to e^+ e^- \gamma$ was used.
The branching fraction was computed in bins of $m_{ee\gamma}$, thus being independent of any assumption on the
$m_{ee\gamma}$ distribution.
Integrating over the single bins in the accessible region gave~\cite{bib:kpieeg}
\begin{equation}
  \Br(\kpmpieeg)_{m_{e e \gamma} > 260 \: \text{MeV}/c^2} 
            = (1.19 \pm 0.12_\text{stat} \pm 0.04_\text{syst}) \times 10^{-8}.
\end{equation}
A single-parameter fit to the $m_{ee\gamma}$ distribution above 260~MeV/$c^2$ gave a value $\hat{c} = 0.90 \pm 0.45$.
Using this value for $\hat{c}$, the total branching ratio was obtained as
\begin{equation}
  \Br(\kpmpieeg) = (1.29 \pm 0.13_\text{exp} \pm 0.03_{\hat{c}}) \times 10^{-8},
\end{equation}
where the last uncertainty reflects the model uncertainty for $m_{ee\gamma}$ below 260~MeV/$c^2$.


\end{document}